**Physics postgraduate teaching assistants' grading approaches: Conflicting goals and practices**

Emily Marshman[1], Alexandru Maries[2], Ryan T. Sayer[3], Charles Henderson[4], Chandralekha Singh[1], and Edit Yerushalmi[5]

[1]*Department of Physics and Astronomy, University of Pittsburgh, 3941 O'Hara St., Pittsburgh, PA 15260*
[2]*Department of Physics, University of Cincinnati, 345 Clifton Ct., Cincinnati, OH 45221*
[3]*Department of Physics, Bemidji State University, 1500 Birchmont Dr. NE, Bemidji, MN 56601*
[4]*Department of Physics, Western Michigan University, 1903 W. Michigan Ave., Kalamazoo, MI, 49008*
[5]*Department of Science Teaching, Weizmann Institute of Science, 234 Herzl St., Rehovot, Israel 7610001*

**Abstract:** Grading can shape students' learning and encourage use of effective problem solving practices. Teaching assistants (TAs) are often responsible for grading student solutions and providing feedback, thus, their perceptions of grading may impact grading practices in the physics classroom. Understanding TAs' perceptions of grading is instrumental for curriculum developers as well as professional development leaders interested in improving grading practices. In order to identify TAs' perceptions of grading, we used a data collection tool designed to elicit TAs' considerations when making grading decisions as well as elicit possible conflicts between their stated goals and actual grading practices. The tool was designed to explicate TAs' attitudes towards research-based grading practices that promote effective problem solving approaches. TAs were first asked to state their goals for grading in general. Then, TAs graded student solutions in a simulated setting while explicating and discussing their underlying considerations. The data collection tool was administered at the beginning of TAs' first postgraduate teaching appointment and again after one semester of teaching experience. We found that almost all of the TAs stated that the purpose of grading was formative, i.e., grading should encourage students to learn from their mistakes as well as inform the instructor of common student difficulties. However, when making grading decisions in a simulated setting, the majority of TAs' grading considerations focused on correctness and they did not assign grades in a way that encourages use of effective problem solving approaches. TAs' perceptions of grading did not change significantly during one semester of teaching experience.

## I. INTRODUCTION

Learning is shaped by grading because grading can communicate to the students (both implicitly and explicitly) the goals and expectations instructors have for their students [1-5]. For example, typical instructional goals of instructors who teach introductory physics [6] are: 1) developing students' understanding of physics concepts and principles [7-12]; and 2) helping students develop systematic problem solving approaches in order to make better use of problem solving as a tool for learning [13-15]. However, grading, as any other instructional decision, is shaped by a vast array of beliefs, goals, knowledge, and action plans triggered by various aspects of the immediate classroom context (e.g., students disagreeing about their grades, peer and administrator expectations, workload, etc.) and some of these considerations might occasionally be conflicting [16]. Thus, it is reasonable to expect that instructors' grading decisions would promote some of their goals, but misalign with other goals.

At large research institutions, postgraduate TAs are frequently the ones responsible for grading students' work. However, many postgraduate TAs are learning to be researchers and instructors concurrently, and they must meet the expectations of both their research advisors and course instructors. The resources accessible to them are usually their own experiences as novice students as well as the requirements of the departments and/or lecturers they assist. TAs usually have a narrow window in time to develop their personal approach towards instruction in general by, for example, clarifying their goals and developing grading practices which adequately transmit their instructional values and beliefs. Here, we present a case-study about postgraduate TAs' goals for grading and their considerations when grading with an eye to inform professional development leaders interested in helping TAs to improve their grading practices.

The study involved 43 first-year postgraduate TAs enrolled in a semester-long TA professional development program at a research university in the US. The data collection tool was designed for professional development to help TAs reflect on their grading practices, and also to facilitate productive discussions around grading that are closely tied to the types of solutions that the TAs themselves are likely to grade in their TA assignments. It probed implicit and, at times, conflicting perceptions by requiring respondents to describe their goals in general and in varying specific contexts. In particular, it made use of several student solutions which had been used in prior studies investigating faculty grading decisions [17]. The solutions were designed to encourage introspection with regard to problem solving approaches that educational literature suggests as a means to promote effective problem solving practices [7-13, 18-20]. To that end, they differed in aspects such as the explication of reasoning or the extent to which they followed a prescribed problem solving approach, i.e., describing the problem in physics terms (e.g., identifying the relevant physics, drawing a sketch, identifying target variable, listing known and unknown variables), planning an approach towards the solution (e.g., articulating sub-problems in terms of sub-goals, explaining how to achieve the sub-goals by means of physics concepts and principles), monitoring the implementation of the solution plan, and evaluating the results. Thus, the data collection tool allowed us to examine the extent to which TAs consider and reward effective approaches to problem solving (i.e., via their grading of student solutions).

TAs worked in groups in which they examined student solutions which portrayed both productive and unproductive approaches to problem solving. The TAs were required to articulate solution features that they graded on as well as their reasons for assigning a particular score. Guidance was provided to help them reflect on their grading decisions, explaining what they valued and what they disapproved of in students' solutions and why. They were also asked whether they would grade differently in a quiz or a homework context. These two contexts were chosen as it was expected that the TAs may have different grading considerations in a quiz vs. a homework. Thus, the activity served to help TAs explicate their goals and grading approaches as well as facilitate a discussion among the TAs to help them examine their instructional decisions in light of their goals.

Along with the literature [21] suggesting that conceptual change follows the realization of a cognitive conflict, it is reasonable to expect that TAs' first encounter with the data collection tool designed to encourage introspection would cause them to become aware of existing inconsistencies, thus causing a cognitive conflict which may motivate changes in their grading decisions. However, regardless of whether changes occurred, instructional approaches could have become more or less ingrained as TAs gained teaching experience. Therefore, we examined TAs' grading decisions and considerations at two instants of time: 1) at the beginning of their teaching career; and 2) after teaching for one semester. This paper addresses the following research questions:

*1: What grading decisions do TAs have at the beginning of their teaching appointment?*

*2: What considerations underlie TAs' grading decisions at the beginning of their teaching appointment?*

*3: To what extent are TAs' grading decisions aligned with their general beliefs about the purposes of grading?*

*4: How do TAs' grading decisions and considerations change after a short professional development intervention and a semester of teaching experience?*

## II. BACKGROUND AND LITERATURE REVIEW

Our study is based on two lines of research: 1) Promoting effective problem solving practices via grading, and 2) TAs' beliefs and practices related to physics teaching and learning and, in particular, the role of problem solving.

### A. Promoting effective problem solving practices via grading

A large body of research [7-12, 22] has investigated differences between experts (successful problem solvers) and novices in problem solving. Both use strategies to help them identify the appropriate actions needed to help bridge the gap between the problem goal and their current state of the solution. However, the types of strategies novices and experts use differ. Novices tend to be disorganized, and typically start by searching for equations that match the provided information and plug in numbers until they get a numerical answer [7]. Novices also often use their intuitive knowledge instead of using formal physics principles [23]. They also attempt to pattern match, i.e., they superficially match a problem solution they had seen in the textbook or class with the problem they are solving primarily focusing on surface features, even if they do not conceptually understand the previously solved problem [23]. In contrast, experts devote significant time and effort in qualitatively describing the problem situation, identifying useful theoretical principles and concepts, and make use of effective representations based on their better organized domain knowledge [11,12]. In addition, experts frequently work backwards, and engage in planning a solution strategy by identifying intermediate goals and means to achieve them [11]. Experts also differ from novices in that they use representations to conceptually analyze and explore problems (even when they are not sure how to proceed) [11]. Experts are also much more adept at self-monitoring, and if necessary, they re-evaluate former steps and revise their choices [9,22,24]. Experts utilize problem solving as a learning opportunity more effectively by engaging in self-repair - identifying and attempting to resolve conflicts between their own mental model and the scientific model conveyed by peers' solutions or worked-out examples [14,15,18-20].

Students' expectations and epistemological beliefs regarding problem solving in physics play a key role in how they approach problems. When students were asked to write a reflective statement at the beginning of a physics course about how they approach physics problems and the methods they use, over 50% of the students stated that they select equations by matching them to the list of knowns and unknowns [25]. Less than 20% of the students claimed that they listed unknowns, drew a diagram, thought first about the concepts involved in the problem, analyzed the problem qualitatively (e.g., identified constraints), and identified sub-problems [25]. Students' epistemological beliefs were found to predict performance. Students who believed that physics knowledge consists of isolated pieces of information that should be remembered performed lower in a one-week project involving an advanced classical mechanics problem than those who did not [26]. In addition, students who reported that they did not elaborate on the links between formulae and theories while solving physics problems had lower scores on laboratory reports [26]. Studies of students in intermediate-level physics courses showed that even they can often get "stuck" in an epistemological frame, i.e., the students' perception as to what tools and skills are appropriate to bring to bear in solving a problem [27]. For example, when reasoning about a physics problem, students sometimes invoke authority and have difficulty switching from that epistemological frame to a more coherent one (e.g., mapping a mathematical representation to a physical system) [27].

To prepare students for future learning [28], students benefit from instruction that fosters both the development of conceptual understanding and promotes effective problem solving practices. Students who solve a large number of problems may become routine experts – they can solve similar problems faster and more accurately but without constructing their conceptual knowledge [29]. As a result, they lack the flexibility and adaptability to solve novel problems. On the other hand, if students are given opportunities to develop both their conceptual understanding and engage in problem solving practices such as qualitatively describing the problem situation using effective representations, planning a strategy for constructing a solution, and self-monitoring, they may become adaptive experts who have a well-organized knowledge structure and engage in effective problem solving practices [29]. This combination of conceptual understanding and effective problem solving may allow students to transfer their learning to solve novel problems [28]. To help students become adaptive experts, students can be given many different types of analogous problems (i.e., problems that require the same physics principles to solve them but have different contexts) and rewarded for explaining the physics principles used, justifying why the principle can be used in a particular situation, and using effective problem solving approaches when solving problems.

Grading is a form of formative assessment [30,31], and provides feedback that moves learning forward, by communicating to learners what practices are useful in learning the discipline [31] and what to focus on in future learning activities [31-36]. Also, grading can promote the use of effective problem solving practices discussed above, for example, by rewarding effective problem solving strategies such as drawing a diagram, listing known and unknown quantities, clarifying considerations in setting up sub-problems, and evaluating the answer. Grading can encourage students to explain their reasoning by placing the "burden of proof" on the student (i.e., requiring the student to explain the reasoning underlying his solution and calculations) and thus provide them with an artifact to reflect on and learn from problem solving (i.e., their own clearly articulated solution in which reasoning and evaluation are explicated) [37].

Docktor and Heller [38] designed a grading rubric along these lines to help students recognize that problem solving is a process requiring use of both content knowledge and effective problem solving strategies. These rubrics assess students' difficulties (in both their content knowledge and problem solving) and include the processes of organizing problem information into a useful description, selecting appropriate physics principles, applying physics to the specific conditions in the problem, using mathematical procedures appropriately, and the overall communication of an organized reasoning pattern [38]. While their rubric also allows for graders to omit grading on a process if it was not required (as judged by the grader), they state that "it is important to consider only what is written and avoid the tendency to assume missing or unclear thought process are correct [38]." This aligns with the concept that instructors should place the burden of proof of understanding on the student and value a logical, coherent solution by grading on explication of reasoning.

### B. TAs' instructional beliefs and practices

Past experiences influence teachers' and TAs' beliefs regarding teaching and learning [39,40]. TAs' past experiences as students shape their intuitive perceptions about teaching and learning, and these views are often highly resistant to change [41-43]. Since postgraduate TAs were recently students in introductory physics classrooms, it is reasonable to expect that TAs' beliefs regarding teaching problem solving are influenced by their prior physics instructors. Hora et al. [44] investigated the beliefs of 56 math and science instructors at undergraduate universities regarding student learning in general, and in particular, in the context of solving problems. Instructors stated that students learn, e.g., by practice and perseverance, articulation of their own ideas and problem solving processes to others, active construction, repetition, and memorization. Some instructors stated that learning occurs over time through sustained engagement with the material. Also, a central goal of instruction for many physics instructors is to improve students' problem solving approaches [6]. Instructors often state that they believe students can learn effective problem solving approaches by watching experts solve problems or reading example solutions, extracting the strategies underlying these solutions, and reflectively attempting to work problems [6]. In regards to grading problem solutions, a study by Henderson et al. [37] demonstrated that most instructors know that there are advantages for students to show their reasoning in problem solutions because it 1) helps students rehearse and improve their problem solving practices and understanding of physics concepts; and 2) allows the instructor to observe and diagnose student difficulties. However, less than half of the instructors interviewed gave students a grade incentive for explaining their reasoning. Many instructors placed the "burden of proof" of student understanding on themselves when assigning a score to a student solution.

Furthermore, the learning experience TAs had and the types of learning environments they believe to have been beneficial for them may influence their beliefs about teaching and learning. For example, in a laboratory context, TAs expect that students learn similarly to them and implement instructional strategies that they perceive were effective for them (but not necessarily beneficial for students) [42]. TAs may acknowledge instructional strategies from educational research, but sometimes disregard them for their own views of appropriate instruction, e.g., that the TA should make the material clear and that students need direct instruction and extensive practice to learn the required concepts [42]. Similarly, TAs' own approaches to problem solving can serve as indicators of their beliefs regarding learning and teaching problem solving. Mason and Singh found that while nearly 90% of postgraduate students reported that they explicitly think about the underlying concepts when solving introductory physics problems, approximately 30% of them stated that solving introductory physics problems merely requires a "plug and chug" strategy [45]. A possible explanation for these findings is that postgraduate students do not engage in problem solving when they solve introductory problems, because these problems are merely exercises for them (i.e., the solution is "obvious" to postgraduate students and they do not need to use effective problem solving approaches that would otherwise be essential). TAs can immediately recognize the principles required to solve a problem and perceive problem solving as not requiring much thought or reflection. Many postgraduate students believed that reflection after problem solving is unnecessary because the problem was so obvious [45]. These findings suggest that TAs who teach recitations or laboratory sections may not model, coach, or assess explication of reasoning or reflection because it was not necessary for them to solve introductory physics problems, although it is highly beneficial to introductory physics students.

Often, postgraduate students receive limited training and feedback related to their TA duties [46] which can result in misalignments between their instructional beliefs and their teaching practices [47-55]. There are often discrepancies between TAs' stated beliefs and their actual classroom practice regarding active participation in the learning process. For example, TAs may endorse the goal of engaging students in sense-making while at the same time, they can devote significant time to transmit knowledge. Also, they state that they value example solutions which reflect effective problem solving strategies but create brief solutions which do not reflect an effective problem solving approach) [51-53]. Lin et al. [53] investigated TAs' beliefs about the learning and teaching of problem solving using example problem solutions. This study revealed a discrepancy between TAs' stated goals and practice. For example, when TAs were asked to evaluate three different versions of example solutions, many valued solutions comprising of features that were supportive of helping students develop effective problem solving strategies. Most TAs expressed process-oriented learning goals (i.e., helping students become more systematic in their problem solving approaches and make better use of problem solving as a tool for learning [56]) when contemplating the use of example solutions in introductory physics. However, their own designed example solutions did not include features supportive of helping students develop of effective problem solving approaches. When TAs were unaware of the conflict between their stated goals and practice, they tended to prefer product-oriented solutions (i.e., solutions in which the reasoning is not explicated [56]). A similar discrepancy may arise in the context of grading, i.e., TAs may have productive beliefs about the role of grading in the learning process, but employ grading practices which do not align with those beliefs.

Since 1) grading is an instructional strategy that can help students learn physics concepts and develop effective problem solving practices; 2) TAs' instructional beliefs and practices are often in conflict; and 3) no study exists on physics TAs' grading pertaining to our research questions, we investigate TAs' goals for grading and their grading practices. This kind of work has potentially strong relations with the extensive phenomenography based "conceptions of learning" and "approaches to learning" research that has previously been conducted as well as the related metacognition/metalearning (i.e., thinking about thinking) that provides insights into the personal awareness and control of knowledge construction [57-61]. In particular, in the research presented here, the postgraduate students are interpreting their goals for grading and their grading practices from their own ontological and epistemological perspectives.

## III. METHODOLOGY AND DATA COLLECTION

### A. Participants

Data were collected from two different cohorts of first-year postgraduate TAs who were enrolled in a professional development program led by one of the authors (C.S.) over two different years. The program was semester-long and was designed to prepare the postgraduate TAs in their first semester of the postgraduate work for their teaching duties and make them reflective educators. The TAs were expected to do approximately one hour of homework each week pertaining to the professional development course, in which various activities took place throughout the semester. Initial activities related to the course focused on general issues related to teaching and learning, e.g., discussion of some physics education research papers on common student difficulties in introductory physics. The discussion of grading practices (described in detail in the data collection section) occurred near the beginning of the semester, followed by discussions of pedagogy, including the use of tutorials and clicker questions as learning tools and the importance of integrating conceptual and quantitative learning. Next, discussions turned to how different problem types (e.g., multiple-choice problems, context-rich problems, problems that are broken into sub-problems, and traditional textbook style problems) can help students learn physics and can be useful in different instructional settings to meet different instructional goals. Towards the end of the semester, the TAs were given a physics problem and asked to present the solution to the TA professional development class as they would in their recitations. These presentations were video-recorded so that the TAs could reflect on their teaching and also receive feedback from other TAs and the instructor. Thus, the grading activity was one of a number of activities all aimed at improving the TAs' effectiveness in and out of the classroom.

In total, 43 first-year TAs were enrolled in the program in the two years (25 in the first cohort and 18 in the second cohort). All the TAs had completed a bachelor's degree in physics or engineering and were admitted to the doctoral program in physics at a research university in the U.S. During their orientation, the TAs attended a university-wide teaching assistant workshop, but this workshop included no discussions of discipline-specific issues in teaching physics. The majority of the TAs were concurrently teaching recitations for introductory physics courses for the first time. During the recitations, the majority of the TAs answered student questions about the homework and solved problems on the board. A few other TAs were also assigned to facilitate a laboratory section or grade students' work in various physics courses for the first time. All the labs were primarily traditional in that they prescribed the set ups to the experiments, the data to collect, the analysis to conduct, etc. A majority of the TAs were also tutors in a physics exploration center where they assisted introductory students with their physics homework and laboratory reports.

The participants' national backgrounds varied; in total, there were 14 postgraduate students from the United States, 17 postgraduate students from China, and 12 students from other countries. There were 5 female TAs and 38 male TAs. The demographics of the TAs in this professional development program are somewhat similar to national norms (see, e.g., ref [62]). Also, all the participants were informed at the end of the semester that some of the data collected from the professional development course would be used for a research study and all of them agreed that their data could be used for this purpose.

### B. Data Collection

#### a. Development and validation of the data collection tool

The data on TAs' goals for grading and grading approaches were collected using a group administered interactive questionnaire (GAIQ) which was previously developed and validated for use with TAs/instructors [63]. The GAIQ is comprised of a series of activities which use worksheets designed to clarify a TA/instructor's ideas about helping students learn physics content and desired problem solving practices. The GAIQ worksheets and artifacts encourage reflection on various aspects of teaching physics problem solving: Designing problems on a particular physics topic with features effective for use in different situations (e.g., questions for peer and class discussions, homework, quizzes, exams, collaborative learning, etc.), designing solutions to homework problems that will help students develop effective problem solving strategies, and grading student solutions. Questionnaires on each aspect of teaching problem solving (e.g., problem types, instructors' example solutions, or grading) involve three stages: 1. TAs/Instructors are individually asked to solve a core problem (Fig 1) suitable for distributing to their students and complete a worksheet eliciting TAs/Instructors' initial ideas about teaching problem solving; 2. TAs/instructors work in groups of three to answer the same questions as in the pre-class worksheet and then each group shares their ideas in a whole class discussion; 3. TAs/instructors individually complete another worksheet in which they can modify their previous answers and connect their ideas to a list of pre-defined features about teaching problem solving developed by the researchers. As mentioned earlier, the GAIQ (data collection tool) was designed for professional development to help TAs reflect on their grading practices. However, it is designed in a way to provide feedback to professional development leaders/researchers and serves as a data collection tool on TAs approaches and attitudes related to grading student work. This means that the GAIQ had a dual role: both as a tool to help the TAs reflect upon and develop productive attitudes related to grading and also as a data collection tool which helped the researchers investigate TAs' grading approaches.

The initial versions of the worksheets were designed using qualitative data from semi-structured interviews with faculty members based on an "artifact comparison" approach [17]. In these interviews, faculty members were asked to make judgments about instructional artifacts which were similar to those they often use in their classes. The artifacts presented to the instructors during interviews were designed to reflect those that would be familiar to physics instructors. In particular, the three types of instructional artifacts were instructors' example solutions, student solutions, and problem types (e.g., problems in multiple-choice format, context-rich form, divided into sub-problems, with and without diagrams, etc.). The artifacts presented to the instructors during interviews were designed to create a context which would activate beliefs that could influence decisions when they select instructional material or pedagogical techniques while teaching [17]. All of the GAIQ activities about instructors' solutions, student solutions, and problem types relate to a single Core Problem shown in Fig. 1 [17]. The core problem was designed, validated and approved by four physics instructors who taught introductory physics courses at the University of Minnesota and was used on final exams. The core problem was also sent to several other instructors of physics courses who reported that the difficulty of the problem was such that it required an average student to use an exploratory decision making process as opposed to an algorithmic procedure [17]. The problem involves synthesis of several fundamental physics concepts and principles. The problem included several features of a context-rich problem [17] (i.e., it was set in a realistic context, was not broken into parts, did not include a diagram, etc.) and is rich enough to allow for interesting variations in students' solutions. Students could potentially solve the problem in different ways. Thus, the problem allows for a spectrum of more or less effective problem solving practices. The student final exam solutions were available, providing a source of authentic student solutions which were used both in Ref. [17] and in the present study. The specific artifacts involving grading activities included five student solutions (see an example of two student solutions in Fig. 2), which were based upon actual students' common answers in the final exam. The artifacts were chosen to reflect differences between expert and novice problem solving from the research literature such as including a diagram describing the problem, explication of sub-problems, justification of solution steps, evaluation of final answer, explication of the scientific principles used, evidence of reflective practices, etc. [17].

Instructors' responses to interview questions about the instructional artifacts revolving around the core problem and five student solutions were used to create the initial GAIQ worksheets including the worksheets on grading [63]. We note that the GAIQ is meant to take the place of individual TA/instructor interviews about the teaching and learning of problem solving. While the development and validation of the GAIQ was a very time-consuming process [17], the GAIQ requires significantly less time than interviews for data collection and analysis. Equally important, it avoids researcher intervention in the process of clarifying the interviewees' responses, and the inter-rater agreement on the coding of the data obtained and interpretation of the data is excellent. Thus, the GAIQ worksheets can be used by researchers and professional developers at different institutions to collect and analyze data and data across different institutions can readily be compared with relative objectivity.

You are whirling a stone tied to the end of a string around in a vertical circle having a radius of 65 cm. You wish to whirl the stone fast enough so that when it is released at the point where the stone is moving directly upward it will rise to a maximum height of 23 meters above the lowest point in the circle. In order to do this, what force will you have to exert on the string when the stone passes through its lowest point one-quarter turn before release? Assume that by the time you have gotten the stone going and it makes its final turn around the circle, you are holding the end of the string at a fixed position. Assume also that air resistance can be neglected. The stone weighs 18 N.

The correct answer is 1292 N.

**FIGURE 1.** Core problem

The initial version of the GAIQ was iterated between the researchers and physics instructors and modified to a version which was administered in the context of professional development for Israeli pre-service and in-service teachers multiple times [64]. After each initial implementation and feedback from the teachers, the GAIQ was refined further until a version was developed that the researchers were satisfied with. The GAIQ tool was then adapted for a professional development program for physics teaching assistants in the United States. The TA professional development program in this study followed suggestions from prominent teacher educators to anchor professional development in collaborative reflection with peer instructors on classroom experiences [65-68]. Reflection on practice serves to enrich instructors' interpretations of classroom experiences, widen the inventory of possible actions instructors might use, clarify instructional goals and examine practice in view of these goals, and provide motivation for the adoption of new instructional strategies. Following these suggestions, the activities in the TA professional development program elicited TAs' initial ideas on different aspects of teaching and learning. Then, the instructor facilitated peer discussions about TAs' ideas, led class discussions in which the instructor provided ideas for "best practices", and also provided opportunities for TAs to reflect on their ideas (for example, opportunities to think about discrepancies in their beliefs about teaching and learning and reflect on changes in their initial ideas).

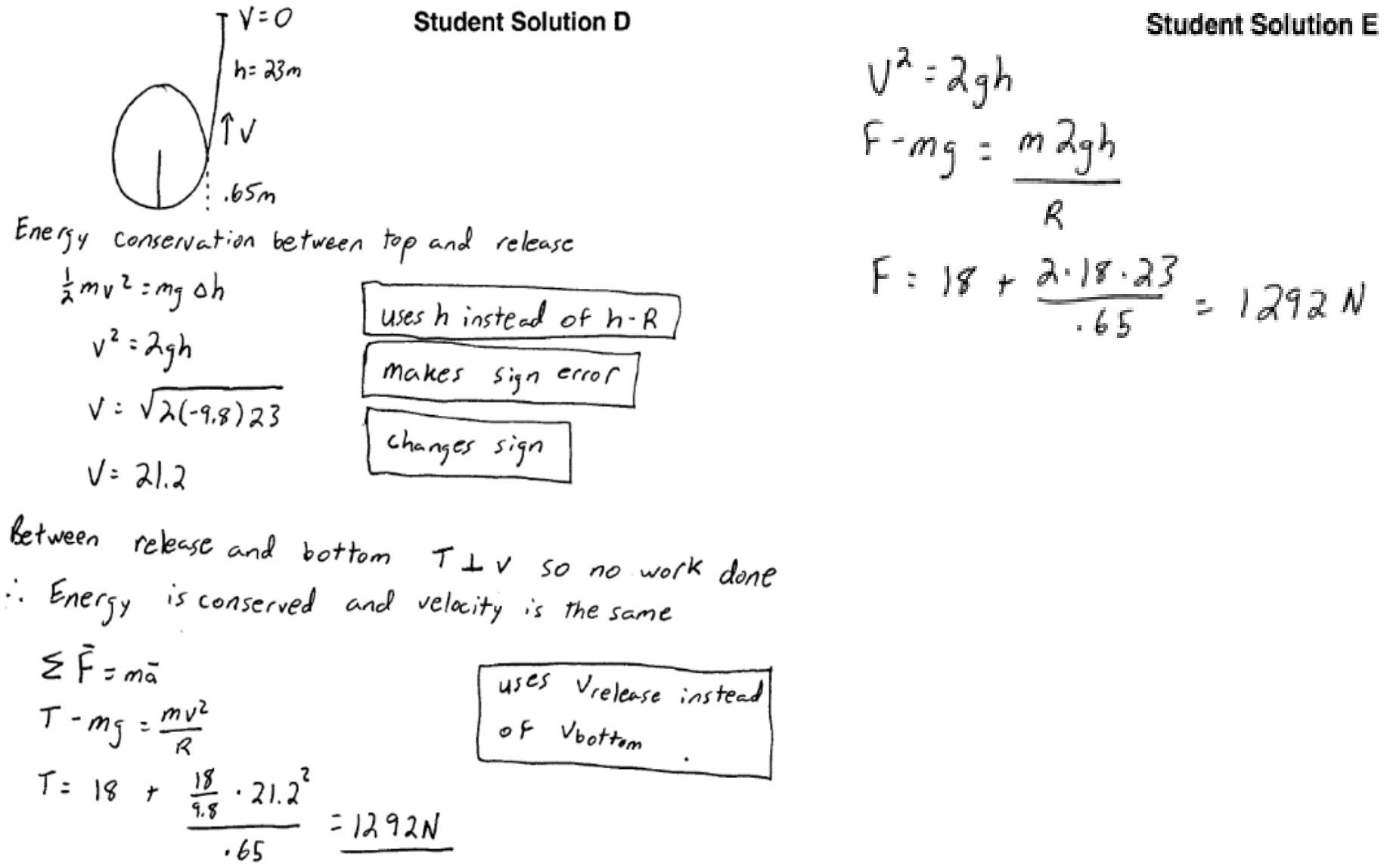

**FIGURE 2.** Student Solution D (SSD) and Student Solution E (SSE)**.**

The GAIQ including the grading activities were implemented in two different semesters of a TA professional development program in the US, and after each implementation, the researchers iterated the version several times between them. A postgraduate student researcher in PER observed the three semesters of the TA professional development program when TAs' worked on the GAIQ. The postgraduate student researcher and two of the authors (E.Y. and C.S.) revised and iterated the GAIQ based upon the TAs' comments and responses. This validation process in the context of the TA professional development program ensured that TAs interpreted all components of the GAIQ appropriately as the researchers had intended.

The artifacts about grading have also been used as the basis of a previous paper on faculty members' grading practices [37]. In this study [37], faculty members were asked to solve the core problem (see Fig. 1) and compare and make judgments about two student solutions (see Fig. 2) to the core problem. These two solutions were chosen because they engender conflicting instructional considerations in assigning a grade [37]. In this paper, we present findings related to the same two student solutions (see Fig. 2) to enable comparison of the findings of the TAs and faculty members' grading practices and values [37]. We suggest that the readers examine the student solutions (see Fig. 2) and think about how to grade them. We note that incorrect aspects of the solutions are indicated by boxed notes. Both solutions arrive at the correct answer. SSD includes a diagram, articulates the principles used to find intermediate variables, and provides clear justifications. The elaborated reasoning in SSD reveals two canceling mistakes, involving misreading of the problem situation as well as misuse of energy conservation to imply circular motion with constant speed. On the other hand, SSE is brief with no articulation of reasoning, and it does not give away any evidence for mistaken ideas. However, the three lines of work in SSE are also present in SSD, suggesting that Student E could have been guided by a similar thought process as Student D.

#### b. Implementation of the data collection tool within the TA professional development program

The TA professional development program consisted of two-hour meetings held weekly throughout the fall semester. Three consecutive weekly sessions at the outset of the program revolved around the GAIQ involving grading activities. Table I shows the sequence of grading activities. The activities served as a learning experience within the professional development program as well as a data collection tool in order to study TAs' grading decisions and considerations in a simulated environment [54,55]. Data regarding grading was collected twice, at the beginning (the second and third weeks) and end of the semester (the last class). Thus, prior to the first data collection, the TAs had only minimal teaching experience in recitations and labs.

The GAIQ sequence and particular worksheet questions are shown in Table 1. The TAs completed a pre-lesson stage, in which they responded to the following questions in the form of an essay:

1) What, in your view, is the purpose of grading students' work?
2) What would you like students to do with the graded solutions returned to them?
3) What do you think most of them actually do?
4) Are there other situations besides the final exam and quizzes in which students should be graded?
5) Does grading serve the same purposes for these situations?

In the pre-lesson stage of the GAIQ, TAs individually graded the student solutions for both homework (HW) and quiz contexts out of a total score of ten points, listed characteristic solution features, and explained why they weighed the different features to obtain a final score. The TAs were told to assume that 1) they were the instructors of the course and thus have ultimate freedom in structuring their grading approaches; 2) they had the authority to make grading decisions; and, 3) their students were aware of how they would be graded. An example response (transcribed) is shown in Figure 3.

During the in-lesson stage of the GAIQ (see Table 1), the TAs worked in groups of three in which they discussed the student solutions and attempted to reach an agreement regarding how to grade them. Afterwards, a representative from each group shared their grading approaches with the entire class. Two of the authors (C.S. and E.M.) were present in the class. C.S. coordinated the class work and led a discussion at the end of the class which highlighted grading "best practices"—i.e., grading approaches that promote effective problem solving. The instructor of the TA professional development program also noted the disadvantages of grading which focused exclusively on correctness. E.M. observed and documented the TAs' comments during the group and whole-class discussions.

The post-lesson stage of the GAIQ (see Table 1) examined the effect of the individual activities, as well as the group and class discussion on TAs' perceptions about grading. The TAs completed an individual worksheet in which they related their initial features to twenty solution features that were defined based on the analysis of TAs' answers to the pre-lesson activity. They were also asked to consider changes in their consideration of features in re-grading the student solutions. This stage intended to allow TAs to rethink their choices as well as to allow the respondents to take part in the categorization of their responses regarding features by using a predetermined set of features.

**TABLE I.** GAIQ sequence of grading activities. The three activities from the beginning of the semester were part of the GAIQ. At the end of the semester, the TAs repeated the first part of the GAIQ (writing an essay about the purpose of grading, completed a worksheet grading student solutions) and completed a reflection activity related to how their grading practices have changed.

| Time | | Activity |
|---|---|---|
| Beginning of semester (GAIQ sequence) | Pre-lesson Individually | 1. TAs wrote an essay regarding the purpose of grading.<br>2. TAs completed a worksheet which asked them to grade student solutions (see Fig. 2) in homework (HW) and quiz contexts, list features of each solution, and explain why they weighed the features to arrive at a final score (see Fig. 3). |
| | In-lesson Groups of 3 | 1. TAs graded the student solutions using a group worksheet.<br>2. TAs participated in a whole-class discussion in which the groups shared their grading approaches and the instructor highlighted grading approaches that promote desired problem solving approaches. |
| | Post-lesson Individually | 1. TAs were given a list of 20 solution features and asked to match those to their initial features and rate how much they liked each feature.<br>2. TAs re-graded the student solutions, keeping in mind the in-class discussions and 20 features they rated. |
| End of Semester | In-lesson Individually | 1. TAs wrote an essay regarding the purpose of grading.<br>2. TAs completed a worksheet which asked them to grade the student solutions (see Fig. 2) in HW and quiz contexts, list features of each solution, and explain why they weighed the features to arrive at a final score (see Fig. 3).<br>3. TAs were given copies of their pre-lesson activities from the beginning of the semester and were asked to reflect on the changes between their responses on the beginning of the semester pre-lesson activities and the end of semester grading activities. |

All components of the GAIQ sequence shown in Table 1 were completed by the TAs within the first month of the TA professional development program when the TAs had very little teaching experience. The end of semester task (see Table 1) was administered to the second cohort of 18 TAs in the last class of the TA professional development program. The in-lesson stage of the end of semester task included the same essay and grading activity as in the beginning of semester pre-lesson stage. In addition, TAs were given copies of their pre-lesson activities from the beginning of the semester and were asked to reflect on how their grading approaches evolved throughout the semester.

| Features: Solution E | Score | | Reasons: explain your reasoning for weighing the different features to result with the score you arrived at. |
|---|---|---|---|
| | Q | HW | |
| No word explanation<br>No figure<br>No error<br>Precise and concise | 10 | 9 | There are no explanations in this solution, which means I could not know whether the student really knows the process or he/she just misdid like solution D. This is why I put 1 point off from this solution if this was HW. However, in the quiz time is limited, I will give a full grade to this solution |

**FIGURE 3**. One component of a sample TA's worksheet (transcribed) related to SSE which was part of the pre-grading activity.

## IV. DATA ANALYSIS AND FINDINGS

***Research question 1: What grading decisions do TAs have at the beginning of their teaching appointment?***

To promote effective problem solving practices, grading should reward explication of reasoning and instructors should place the burden of proof of on the student [37]. Thus, while SSD would lose points for incorrect physics, he/she would receive points for explaining his/her reasoning [37]. However, SSE would receive a lower score because this solution does provide reasoning and it is impossible to know if the student applied physics principles correctly. The instructor places the burden of proof of understanding on the student by deducting points for not explicitly showing evidence of his/her thinking and understanding in SSE [37].

The grading decisions made by TAs (i.e., their scoring of student solutions) for SSD and SSE are represented on a graph of which shows their SSE and SSD scores in a quiz as well as in a homework context (see Fig. 4). The smallest bubbles represent one TA, and larger bubbles represent multiple TAs (the number of TAs at a particular point is related to the size of the bubble and corresponds to 1, 2, 3, or 4 TAs). TAs who are above the diagonal line in the graphs scored SSE higher than SSD. Few TAs gave SSE and SSD the same score in either the homework or quiz context (represented by few bubbles on the diagonal line).

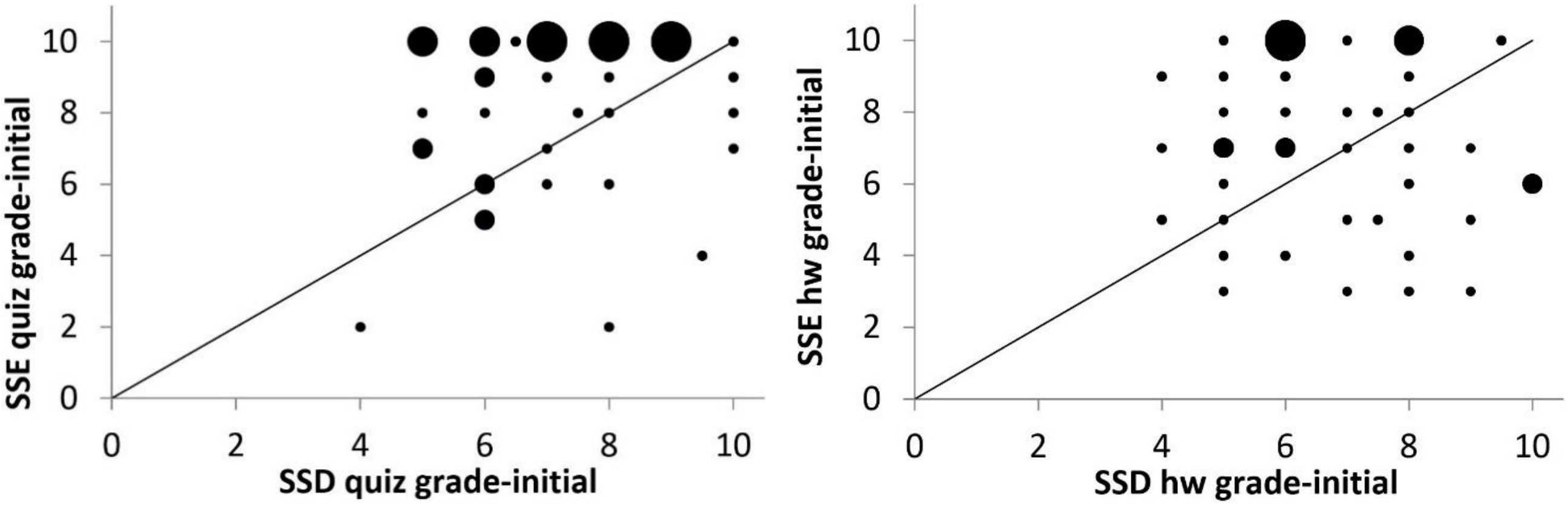

**FIGURE 4.** Distribution of 43 TA grades on SSD and SSE at the beginning of the semester (initial), quiz and HW. The size of the bubble represents the number of TAs at that particular point and correspond to 1, 2, 3, or 4 TAs.

In the quiz context, many more TAs graded SSE higher than SSD ($N$ = 28, 65%) compared to those who graded SSE lower than SSD ($N$ = 10, 23%) and, as a result, the average score for SSE was higher than for SSD ($SSE_{avg}$ = 8.3 compared to $SSD_{avg}$ = 7.1, $p$-value (t-test) = 0.010). We found a similar gap in the homework context, although the gap is somewhat softened: 58% of TAs ($N$ = 25) graded SSE higher than SSD while 35% ($N$ = 15) graded SSE lower than SSD. We also found that in homework, the averages were comparable ($SSE_{avg}$ = 7.1 and $SSD_{avg}$ = 6.7, $p$-value (t-test) = 0.41). We interpret these results to mean that the majority of TAs preferred a solution which lacks reasoning and possibly obscures physics mistakes (SSE) over a solution which shows detailed reasoning but reveals canceling physics mistakes (SSD). Those who rewarded the use of desired problem solving approaches and explicit reasoning in a homework context were more reluctant to do so in a quiz context. This suggests that in a quiz context, TAs are more likely to transmit a message that is counterproductive to promoting these behaviors, i.e., that correctness is more important than explication of reasoning.

***Research question 2: What considerations underlie TAs' grading decisions at the beginning of their teaching appointment?***

We determined TAs' considerations in grading by analyzing: 1) the solution features they mentioned and graded on; and 2) the reasons they stated for assigning a final score. We will discuss our methods and analysis of these two components in the following sections.

**1. Solution features mentioned/graded on**

*Analysis.* The pre-lesson stage of the GAIQ sequence asked TAs to grade student solutions SSD and SSE, list solution features, and explain their reasons for why they weighed the different features to arrive at a final score (see Fig. 3). Data analysis involved coding the features listed by TAs in the worksheets into a combination of theory-driven and emergent categories. Twenty-one solution features were identified. We made a distinction between features that were merely mentioned or weighed in grading. For example, in Fig. 3, the sample TA listed "no figure" as a feature in SSE (solution feature "figure" was considered "mentioned"), but when assigning a grade, s/he did not refer to this feature when explaining how s/he obtained a score (solution feature "figure" was not included in grading). Thus, the sample TA would be counted as mentioning solution feature "figure" but not counted as grading on it. A TA who graded on a solution feature was counted as both mentioning and grading on it. For example, if the sample TA had not written "no word explanation" in the Feature column, the feature "explanation" would have still been considered to be mentioned because this TA wrote "There are no explanations in this solution" as a reason for why he weighed the feature to arrive at a final score he/she would assign to this solution (the feature would also be considered to be graded on). The coding was done by four of the researchers. In cases where disagreement occurred, this was usually due to vagueness in the wording of TAs' written statements. The researchers made use of the TAs' answers in the post-lesson stage (see Table 1) to clarify vague statements made by TAs. After comparing codes, the researchers discussed any disagreements during multiple meetings until agreement better than 90% was reached.

To help interpret the data collected, the features were grouped into the 5 clusters shown in Table II. Each solution feature listed by a TA was entered into only one cluster. Cluster 1 (C1) includes features related to promoting effective problem solving practices [7-13] (i.e., initial problem analysis as well as evaluation of the final result). Cluster C2 involves features related to explication of reasoning (i.e., articulation and justification of principles). Features in C2 are written statements in the student solution that clearly invoke principles used to solve the problem and justify why the principles were used. Cluster 3 (C3) includes domain-specific features, such as invoking relevant physics principles and applying them correctly. Cluster 4 (C4) includes features related to elaboration which emerged during the coding process, e.g., "written statements," "good presentation," "solution in steps," and "conciseness." These features were not assigned to the "explication" category C2 because they were unclear. Cluster C2 is focused on the explication and justification of the physics principles, whereas C4 is more about general communication of the solution. For example, we could not differentiate whether a TA who wrote "written statements" meant that the student solution includes an explicit statement of a principle in writing, explicit justification of a principle in writing, or simply a written statement. Thus, we coded "written statements" as belonging in the general category C4 elaboration. Similarly, if a TA says that a solution is "organized" he/she could mean that the solution is neatly written or that it is systematic. Cluster C4 also involves solution features related to lack of elaboration, e.g., conciseness (this feature was mentioned most often by TAs when they graded the brief student solution SSE). Finally, Cluster 5 (C5) focuses on correctness of algebra and final answer.

TAs' grading decisions were somewhat different in a quiz and a homework context, but we found little difference in the feature clusters they considered in these contexts. Thus, the findings presented relate to grading considerations for the quiz context (the percentages of the TAs who mentioned and graded on clusters in the homework are shown in the appendix, Fig. A1).

**TABLE II**. Sample features sorted into clusters and sample citations

| C1<br>Problem description<br>& evaluation | Visual representation (e.g., "diagram," "figure," "graph"); articulating the target variables and known quantities (e.g., "knowns/unknowns," "list of variables," "nothing labeled"); evaluation of the reasonability of the final answer (e.g., "check," "double check what they did") |
|---|---|
| C2<br>Explication and justification of problem-solving approach | articulation of principles (e.g., "labels energy conservation use," "text showing knowledge of concepts"); justifying principles (e.g., "justify steps," "explanation of why he uses $velocity_{release}$ instead of $velocity_{bottom}$," "explained the reason he used the formulas," "explanation for constant velocity," "no demonstration for why the first equation holds") |
| C3<br>Domain knowledge | Essential principle invoked (e.g., "sums forces, energy conservation," "Newton's 2nd law, conservation of energy," ") ; essential principle is applied adequately ("made some mistakes on applying the velocity," "uses release velocity instead of the velocity at the bottom," "wrong height," "don't know how to apply basic concept correctly," "conceptual errors") |
| C4<br>Elaboration | C4.1 Explanation; written statements (e.g., "verbal explanations," "narration," "word explanation," "no text," "doesn't explain anything," "no words" ) |
| | C4.2 Organization (e.g., "good presentation," "well-organized," "no structure"); showing algebraic steps (e.g., "solution in steps") |
| | C4.3 Conciseness (e.g., "short and concise," "short and sweet") |
| C5<br>Correctness | Algebraic errors (e.g., "makes sign error," "algebra mistakes," "some math errors"); correct final answer (e.g., "final result right," "correct answer") |

*Findings* (see Fig. 5). We found that for clusters C1 and C2, many more TAs mention these features than actually use them when assigning a grade. In particular, roughly 50% of the TAs mentioned features from the problem description and evaluation cluster (C1), but less than 20% considered these features in grading, regardless of whether they were present (as in SSD) or missing (as in SSE). Similarly, a larger percentage of TAs mentioned features involving explication of reasoning (cluster C2) than those who graded on explication. However, in contrast to the description and evaluation cluster that was treated similarly whether it was present (SSD) or missing (SSE), a larger portion (26%) of TAs took the explication cluster into account when they graded SSD (where explication was present) than when they graded SSE. Only 14% of the TAs graded on explication in SSE, where it was missing. Similar to TAs' considerations involving C1 and C2, more TAs mentioned features from C4.1 (i.e., explanation, written statements) than graded on these features (see Fig. 5). Very few TAs (~10%) considered explanation and written statements in their grading of SSD, a solution which includes many written statements. Somewhat more TAs (~30%) considered missing explanations and written statements when grading SSE. These data suggest that while TAs may be aware of features related to explication of reasoning and desired problem solving practices, they were not committed to grade on these same features.

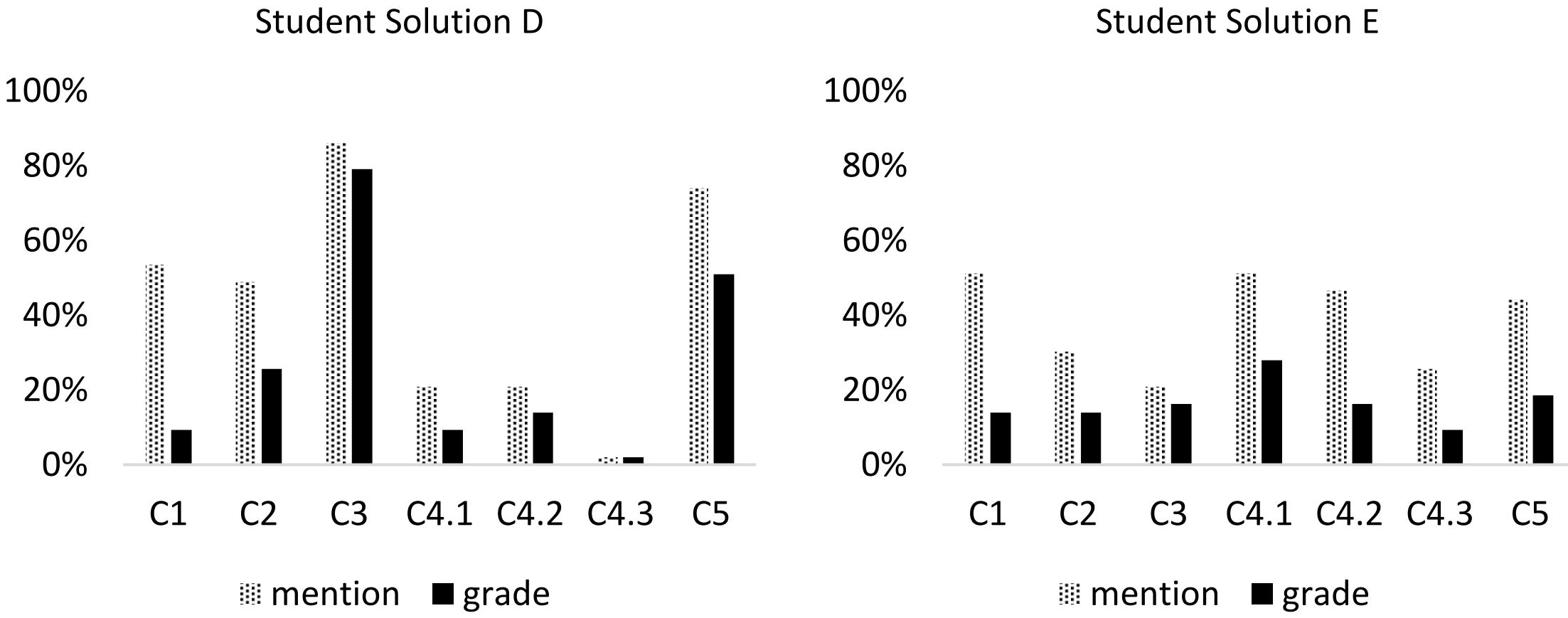

**FIGURE 5**. Percentage of TAs mentioning and grading on features from clusters C1-C5 in SSD and SSE in a quiz context ($N = 43$ TAs).

The cluster most often considered in grading was domain knowledge (C3). Over 80% of all TAs claimed that they grade on features related to domain knowledge (C3) in SSD, deducting points for physics concepts and principles that were inadequately applied. Additionally, approximately 50% of all TAs said they grade on correctness (C5) in SSD, deducting points for explicit algebraic mistakes. Fewer TAs said that they grade on domain knowledge (~30%) or correctness (~ 20%) in SSE, where no apparent mistakes were evident. These data suggest that TAs used a subtractive grading scheme, removing points from SSE for missing explanations (C4.1), but not weighing this cluster in grading SSD, where it is present. Similarly, TAs deducted points from SSD for lacking domain knowledge (C3) and correctness (C5), but did not weigh this cluster when grading SSE, where there are no apparent mistakes.

### 2. Reasons for grading

The data for the analysis of TAs' reasons for grading was collected in several parts of the pre-lesson stage of the GAIQ sequence. First, TAs discussed the purposes for grading in the essay they wrote. TAs were also asked to explain their reasons for weighing different solution features the way they did to arrive at a final score. Open coding was used [69] to generate initial categories grounded in the actual data. The coding of reasons and purposes for grading was completed by four of the researchers. After comparing codes, any disagreements were discussed and the categories were refined until better than 90% agreement was reached.

The reasons for the grade on SSD mostly reiterated the importance of the features included in C3 and C5 (physics knowledge and correctness) and very few TAs mentioned other reasons for scoring SSD in a particular manner. However, many reasons surfaced when TAs graded SSE and these are shown in Table III (we do not show reasons for the grade on SSD since TAs mainly focused on physics knowledge and correctness). TAs' reasons for grading SSE included "adequate evidence" (i.e. whether the solutions allowed the TA to understand the student's thought process), time/stress (i.e., on quiz students do not have time and are too stressed to elaborate their reasoning), and aesthetics (i.e., physics problems should be solved in a brief, condensed manner). Table III shows the number and percentages of TAs (total $N = 43$) who consider these different reasons when grading SSE in the quiz vs. homework contexts. In the quiz context, nine TAs took the burden of proof of student understanding on themselves, stating: "SSE is brief, but I can still understand what was done" and "the student obviously knew what he was doing." In contrast, nine TAs mentioned that SSE contained inadequate evidence in the quiz context, stating that "we cannot determine if he has fully understood the points of the problem" and "it doesn't show if he/she is actually thinking correctly." Six TAs noted that they would be lenient in grading the quiz because of the time limitations in a quiz context. Additionally, five TAs mentioned aesthetics as a reason for the grade on SSE in the quiz context, stating that they liked the conciseness of SSE. In the homework context, a larger number of TAs ($N = 17$) noted that SSE contains inadequate evidence of understanding. Thus, while little difference was found between the solution features mentioned and graded on in the homework and quiz contexts, TAs' consideration of evidence of students' thought processes and consideration of time limitations in a quiz may result in the SSE grade differences in the homework and quiz contexts (i.e., more TAs graded SSE higher in the quiz context as opposed to the homework context).

**TABLE III.** Reasons for SSE grade in the quiz and homework (HW) context, numbers of TAs, and percentages of TAs mentioning reasons. Each TA could provide more than one reason (total $N = 43$ TAs).

| Reason | Sample citation | Quiz $N$ (%) | HW $N$ (%) |
|---|---|---|---|
| Adequate evidence | "SSE is brief, but I can still understand what was done" | 9 (21%) | 6 (14%) |
| Inadequate evidence | "He didn't prove that he understood the problem or accidentally [got it]." | 9 (21%) | 17(40%) |
| Time/stress | "In the quiz in which time is limited, I will give a full grade to this solution" | 6 (14%) | 0 (0%) |
| Aesthetics | "The student had the right idea of how to approach the problem in the simplest way." | 5 (12%) | 4 (9%) |

***Research question 3: To what extent are TAs' grading decisions aligned with their general beliefs about the purposes of grading?***

The goals TAs' have with regards to the purposes of grading (i.e., their answers to the question asked in the pre-lesson stage of the GAIQ, "What, in your view, is the purpose of grading students' work?") fell into four categories: 1) to provide a learning opportunity for the student, 2) to provide instructors with feedback on common difficulties of their students, 3) to provide institutions with grades, 4) and to motivate students (e.g., to turn in their homework or to study harder).

Almost all of the TAs stated that grading serves as a learning opportunity for the student—to reflect on their physics mistakes (content knowledge) and "weaknesses" in problem solving (practices) and learn from them (see Fig. 6). Approximately half of the TAs stated that it is for the benefit of the instructor to understand student difficulties.

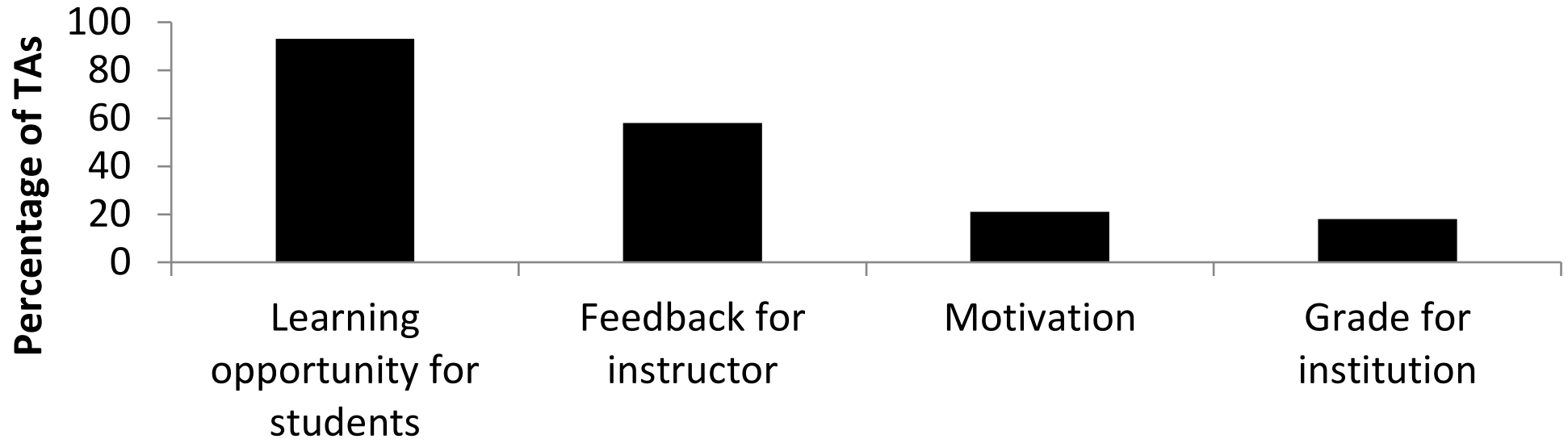

**FIGURE 6**. Responses to the purpose of grading before teaching experience and professional development.

Our findings suggest that most TAs have goals that are aligned with formative assessment goals (i.e., providing feedback to both the student and instructor) as opposed to summative assessment goals (i.e., rank students or assigning a final grade). TAs' grading decisions are aligned with these stated goals in that they provide students with feedback on physics errors that would allow them to learn from their mistakes. However, their reluctance to deduct points where no apparent physics mistakes were made, even when the student did not explain his/her reasoning nor did he/she articulate use of principles, shows that their grading practices are inconsistent with their stated goals for grading. It is also possible that TAs have a narrow perception of formative assessment in which case they may be unaware of the inconsistency, however, this inconsistency is present whether or not the TAs are aware of it. TAs' grading decisions did not encourage students to provide evidence about their thinking that would enable students to reflect and learn from their mistakes. Furthermore, while several TAs stated that students should develop better problem solving practices by reflecting on their graded solutions, few TAs graded on problem solving practices such as problem description and evaluation.

TAs' acceptance of inadequate evidence may also undermine their stated purpose for grading as a means of determining common student difficulties. If a TA takes the burden of proof of student understanding on themselves, filling in gaps in students' missing reasoning, they might miss common difficulties (e.g., student E could be guided by the same incorrect reasoning as student D, but a TA whose grading approach is primarily focused on correctness will assume that student E understands the problem).

***Research Question 4: How do TAs' grading decisions, considerations, and beliefs change after a short professional development intervention and after a semester of teaching experience?***

*Data Collection and analysis.* To examine how the brief professional development class changed TAs' grading beliefs and grading decisions, one of the researchers (E.M.) took notes during group and whole-class discussions (i.e., the in-lesson stage of the GAIQ, see Table 1), which was intended to elicit conflicting viewpoints about grading. To investigate how one semester of teaching experience changed TAs' grading perceptions, the 18 TAs from the second cohort were asked to complete a grading activity at the end of the semester. The final grading activity included the same components as the pre-lesson activity of the GAIQ at the beginning of the semester (i.e., TAs were asked to write an essay regarding grading and grade the student solutions SSD and SSE again). After the TAs wrote the essay and graded the student solutions, they were provided with the worksheets they completed in the pre-lesson stage at the beginning of the semester (see Table 1) to reflect on changes in their grading. The same data analysis as in the pre-lesson activity was completed on the final grading activity. We make comparisons between the initial grading activity completed by 43 TAs and the final grading activity completed by 18 TAs. We note that the responses to the initial grading activity of the 18 TAs from the second cohort were not significantly different than the other 25 TAs' responses to the initial grading activity in the first year of the TA professional development program, so a comparison between the initial grading activity completed by the 43 TAs and the final grading activity completed by the 18 TAs is reasonable. We discuss below the findings in the change in TAs': 1) grading decisions; and 2) grading considerations and beliefs about the purpose of grading.

## 1. Findings - Observation of group discussions within the professional development intervention

In the group discussions, many of the groups scored SSE highly (i.e., gave a score of 9/10 or 10/10). It was often the case that all three of the TAs in one group had previously given SSE a score of 10/10 on their individual worksheet in the pre-lesson stage of the GAIQ. As a result, in the group grading activity, all three TAs agreed on a final score of 10 for SSE. Other groups stated that there were disagreements in their group about the grading of SSE, and they could not come to a consensus. Furthermore, they were unable to suggest ways to resolve this conflict when reporting their group grading to the entire class. In summary, the group discussions did not result in acknowledgment of the roots of disagreements and resolution between different points of view.

## 2. Findings - Change in TAs' grading decisions after one semester of teaching experience

We found that TAs' grading decisions (i.e., how they scored SSE and SSD) remained relatively constant. Similar to the beginning of the semester, at the end of the semester, the majority of the TAs graded SSE significantly higher than SSD (see Fig. 7). TAs who are above the diagonal line in Fig. 7 scored SSE higher than SSD. The gap in average quiz grade between SSE and SSD of the subgroup of 18 TAs became larger over the course of the semester (the average SSE score changes from 7.7 to 8.3 and the average SSD score changes from 7.0 to 6.6 from the beginning to the end of the semester). Thus, the TAs were more likely to focus on correctness as opposed to explication of reasoning in a quiz context when grading at the end of the semester compared to the beginning.

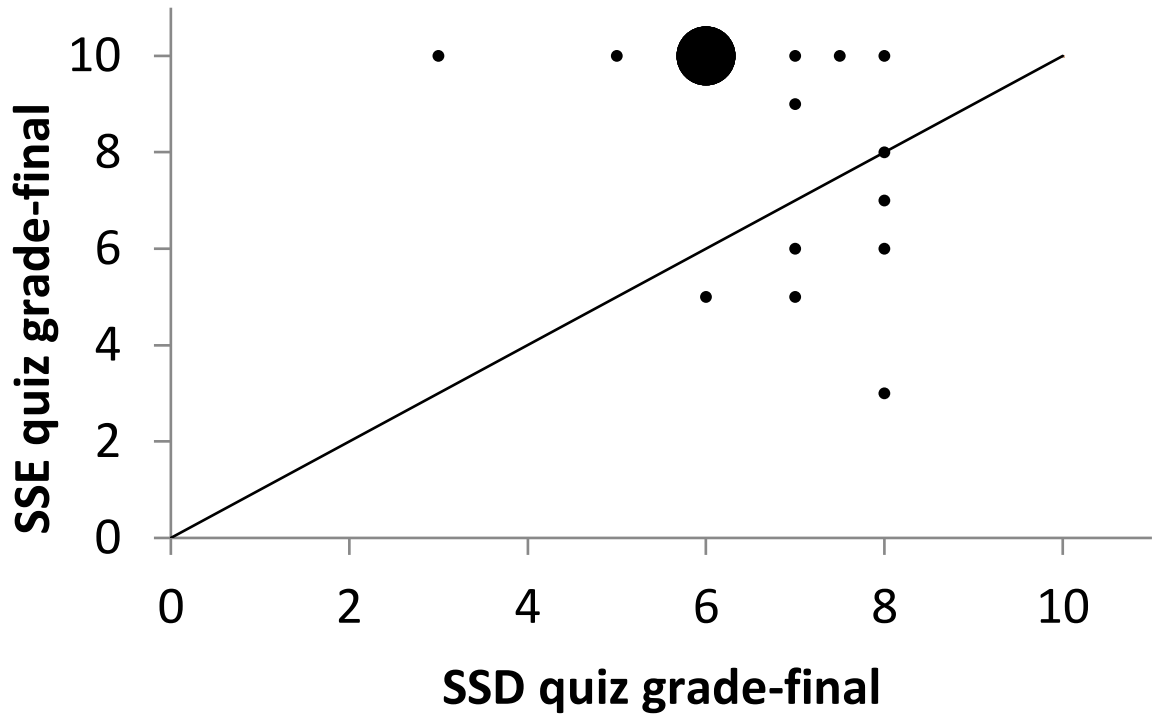

**FIGURE 7**. Distribution of 18 TA quiz grades (SSE vs. SSD) at the end of the semester (final). The size of the bubble represents the number of TAs at that particular point and corresponds to 1 or 4 TAs.

## 3. Findings - Change in TAs' grading considerations and beliefs about the purpose of grading

We also found little change in the distribution of solution features mentioned and graded on by TAs. Regarding reasoning TAs used for assigning a specific grade, the TAs' stated reasons for the final grade on SSE remained approximately the same. We do not show reasons for the grade on SSD since TAs mainly focused on physics knowledge and correctness. Table IV shows that there was a small increase in the percentage of TAs who stated that SSE does not give evidence of understanding in the quiz context after teaching experience and the professional development program (from 21% to 33% of the TAs). However, this did not translate into the TAs grading SSE lower, in fact, their average score on SSE was higher at the end of the semester despite a higher percentage of TAs noting the lack of evidence of understanding in SSE compared to the beginning of the semester.

**TABLE IV.** Reasons for the final grade on SSE in the Quiz and homework (HW) contexts after (Final) teaching experience and PD. TAs could state more than one reason.

| Reasons for the SSE grade | Final (N=18 TAs) | |
|---|---|---|
| | Quiz | HW |
| Adequate evidence | 4 (22%) | 1 (6%) |
| Inadequate evidence | 6 (33%) | 7 (39%) |
| Time/stress | 2 (11%) | 0 (0%) |
| aesthetics | 0 (0%) | 0 (0%) |

Lastly, we found that TAs' general beliefs about the purpose of grading did not change significantly (see Fig. 8). The majority of TAs continued to state that grading is a means for students to learn from their mistakes. However, the percentage of TAs who stated that grading can serve as a formative assessment tool for the instructor decreased by approximately 20%. The number of TAs who stated that grading is a means to give a final grade (i.e., summative assessment tool) increased by approximately 20%. We speculate that their teaching experiences may have partly resulted in this change: while TAs may have initially believed that student solutions provide feedback to the instructor as to what difficulties are common among students, the TAs' grading experiences may have instilled in some TAs the belief that surveying student solutions to determine common difficulties is impractical given the amount of grading required. They may then believe that the purpose of grading is primarily to provide a learning opportunity for students and a means to assign students' grades for the institution. In summary, TAs' grading practices, considerations, and stated purposes for grading did not change significantly after a brief professional development intervention and one semester of teaching experience.

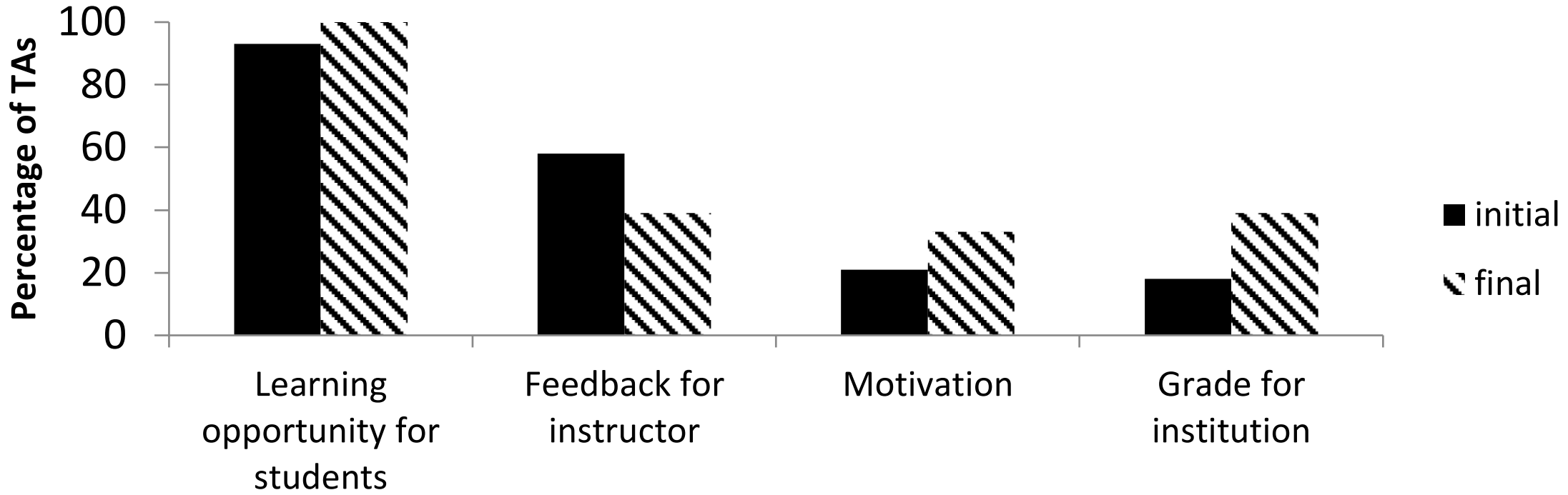

**FIGURE 8**. Responses to the purpose of grading before (initial, $N = 43$ TAs) and after (final, $N = 18$ TAs) teaching experience and professional development.

***Study Limitations***

The findings of this study are contextualized in a task mimicking a "real" grading situation as closely as possible. However, since the TAs graded designed student solutions, the results are valid in this context and shed light on TAs' intended practice rather than TAs' actual grading approaches. In actual practice, TAs' grading approaches may become even more focused on correctness as opposed to encouraging desired problem solving approaches due to external factors, such as a large grading workload and a lack of control and involvement in the design of courses, in particular, homework, quizzes, and exams. In addition, even though TAs were told to grade the student solutions as the instructor of the course (i.e., that they are in control of the class and have told their students how they will be graded), it is possible that TAs' actual grading practices (when they are a TA in another instructor's class) conflicted with how they would like to grade and impacted their responses on the grading activities.

In the TA professional development program, we have also given TAs student solutions to a different context-rich physics problem that is isomorphic to the core problem in this study (see Fig. 1). One student solution was similar to Student Solution D in that it included many effective problem solving practices and showed reasoning and the other student solution was similar to Student Solution E in that it was very brief with only three lines of work. Although we do not discuss the details of this research in the present study, we found that TAs graded the analogous student solutions similarly. This suggests that TAs' grading practices are fairly consistent across analogous types of student solutions and that the way the TAs grade SSD and SSE in this study may provide some evidence about their grading practices in general.

The study was designed to portray TAs' practice and considerations, but further study is needed to learn what factors shape these practice and considerations. One possible factor is TAs' prior educational experiences. The study involves a common mixture of TAs [70] who had experienced undergraduate education in different countries (14 from the U.S., 17 from China, 12 from other countries). Prior research has shown that American, Chinese, and other international TAs perform similarly in identifying common student difficulties [71]. However, grading practices and considerations might be more sensitive to institutional cultures in different countries. We did find small differences between international TAs compared to those of American TAs, however, it is not possible to determine whether differences are significant between the groups due to the small numbers of each group[1]. Additional studies are needed to corroborate the results and draw more robust conclusions about TAs' grading approaches.

## V. SUMMARY AND DISCUSSION

TAs' grading decisions and considerations were examined when they entered their teaching appointment and after one semester of teaching experience and professional development. We found that most TAs perceived the goal of assessment as formative, i.e., helping students learn from their solutions and providing feedback to the instructor about common student difficulties. Some TAs also referred to the role of assessment in encouraging students to develop effective approaches to problem solving. Half of the TAs realized the existence of solution features reflecting effective problem solving approaches when describing students' solutions. However, most of the TAs graded a solution which provides minimal reasoning while possibly obscuring physics mistakes higher than a solution that shows detailed reasoning and includes canceling physics mistakes. This tendency was most evident in a quiz context and somewhat softened in a homework context. When asked to list the features they grade on, TAs commonly did not state that they grade for the solution features representing systematic problem solving, whether in a quiz or homework context. Instead of weighing features representing systematic problem solving in grading, many TAs focused on correctness in domain knowledge, algebraic procedures, and final answer.

The TAs graded on solution features differently in a quiz and homework context, considering differently the extent to which the solution should provide *evidence* that would allow instructors to diagnose students' work. TAs were sometimes conflicted about the issue of evidence in light of other reasons involving time limitations and stress in a quiz. They may have resolved their conflict by softening the requirement for evidence in a quiz context. Our results indicate that many TAs 1) claimed that grading should support the goal of helping students develop desired problem solving approaches and 2) were aware of solution features reinforcing this goal. However, very few TAs graded on these features in the homework and quiz contexts.

Finally, we found that there was little change in TAs' grading practices, considerations, and beliefs after the brief professional development intervention regarding grading and one semester of teaching experience. TAs maintained their general goals for grading – to provide a learning opportunity for the student as well as to provide instructors with feedback on common difficulties of their students. However, TAs' grading decisions and the features they graded on did not change significantly after a semester of teaching experience. In particular, they still did not increase their rewarding of explanations and desired problem solving practices at the end of the semester.

Our findings regarding TAs' grading decisions and considerations are aligned with prior research on physics faculty grading practices [37] in which the instructors often faced internal conflicts when assigning a grade. Most instructors resolved these conflicts by placing the burden of proof on themselves rather than on the student (i.e., they were filling in gaps in student's reasoning in cases where evidence of reasoning was ambiguous) [37]. The results of our study also echo the findings of Lin et al. [53], who found that the goal of helping students develop a systematic problem solving approach underlies many TAs' considerations for the use of example solutions. However, TAs do not use many features described in the research literature as supportive of this goal when designing problem solutions, despite being aware of those features and their importance.

A possible explanation for TAs' grading preferences is their prior experiences as students. Since TAs are recent undergraduate students, it is reasonable to expect their grading approaches would reflect the manner they were graded as undergraduates. The prevailing culture in physics classrooms is determined by the faculty, and as shown by Henderson et al., they often do not give incentives for showing reasoning [37]. Another possible explanation for our findings is that introductory physics problems are essentially exercises for TAs, thus, they do not feel the need to explain their reasoning or reflect on their problem solving process [45] and do not think it appropriate to require their students to do something that they do not find valuable to do themselves.

The short professional development intervention did not serve to trigger in TAs a conflict between various goals and practice and provide tools for aligning their grading practices to better match their goals. These findings are aligned with prior research showing that it is difficult for teachers to alter their views on student learning and that professional development should be long term, allow teachers to bring evidence from the class, and reflect on their practices in light of their goals to allow a meaningful change process [21].

---

[1] While it is difficult to make judgments about the trends, international TAs (non-Chinese) were more likely to score SSD higher than SSE than other groups. On average, they scored SSD higher than other TAs: The average score on SSD (quiz context) for postgraduate TAs from the U.S. is 7.3, from China, 6.4, and from other countries, 7.9. The average score on SSE (quiz context) for TAs from the U.S. is 7.9, from China, 8.9, and from other countries, 7.8. In a quiz context, 82% of the Chinese TAs scored SSE>SSD, 64% of American TAs scored SSE>SSD, and 42% of TAs from other countries scored SSE>SSD.

We conclude that in order for professional development programs to help TAs improve their grading decisions and considerations, an extensive grading intervention is needed that allows sufficient time, structure, and support for TAs to reflect on their actual grading in light of possible goals. The findings of this study could inform professional development providers to design activities that would clarify possible goals for grading and examine them in consideration of the research literature on promoting use of effective problem solving approaches. In particular, examining the manifestation of these goals with respect to various features of systematic problem solving may help TAs improve their grading approaches. Furthermore, professional development could relate to the findings regarding the difference in TAs' grading approaches in a homework and quiz context, in particular, with reference to what TAs perceive as adequate evidence for students' reasoning leading to a solution. Explicating the difference and having TAs discuss the messages sent by quiz grades versus homework grades may assist TAs in improving their grading as an instructional tool that help students learn from problem solving.

## ACKNOWLEDGEMENTS

We thank the members of the physics education research group at the University of Pittsburgh as well as the TAs involved in this study. We thank the National Science Foundation for award DUE-1524575.

## APPENDIX

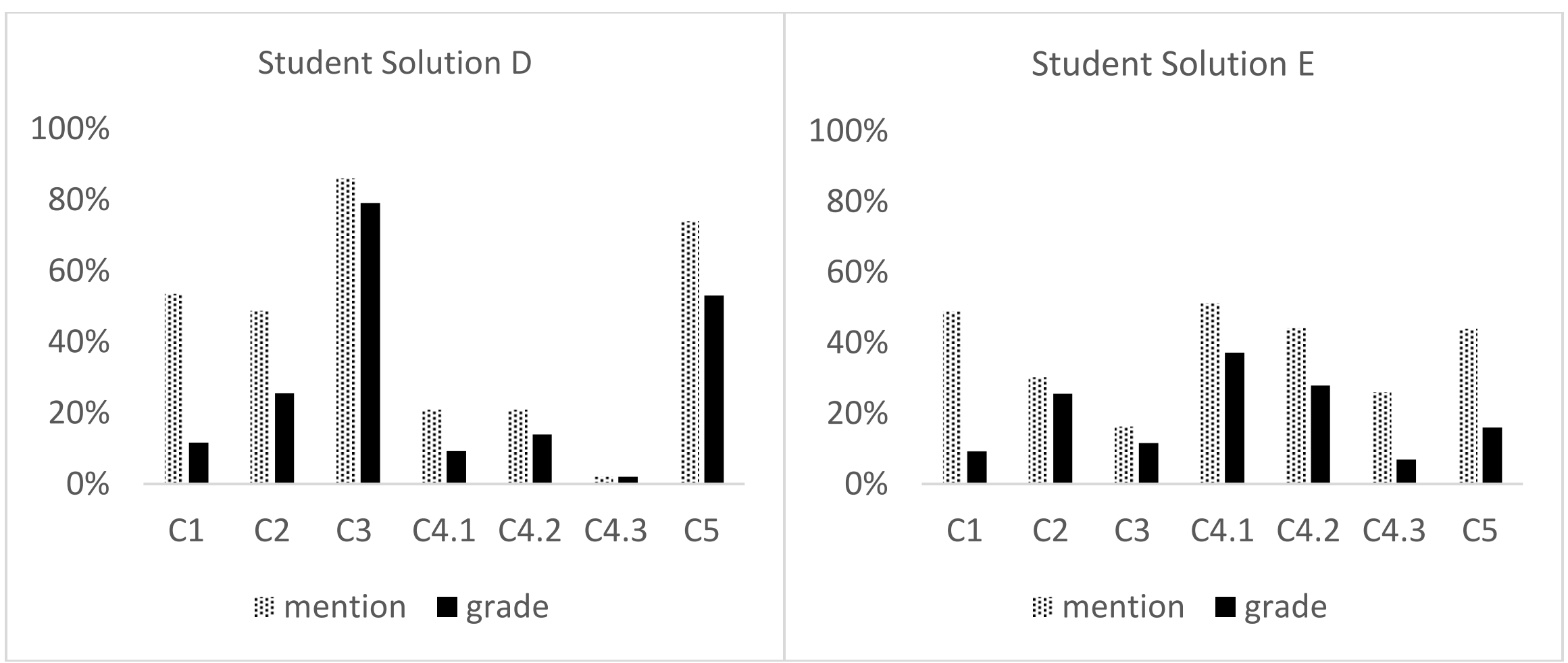

**FIGURE A.1**. Percentage of TAs mentioning and grading on features from clusters C1-C5 in SSD and SSE in a homework context ($N = 43$ TAs).